\title{\huge Borrowing of information across patient subgroups in a basket trial based on distributional discrepancy}

\author{HAIYAN ZHENG$^{1, \ast}$, JAMES M.S. WASON$^{1,2}$ \\[8pt]
\textit{$^1$Biostatistics Research Group, Population Health Sciences Institute, Newcastle University, U.K.} \\
\textit{$^2$MRC Biostatistics Unit, University of Cambridge, U.K.}
\\[8pt]
$^\ast$Email: \href{mailto:haiyan.zheng@newcastle.ac.uk}{haiyan.zheng@newcastle.ac.uk}
}

\date{ }

\documentclass[11pt]{article}

\usepackage[margin=0.78in]{geometry}
\usepackage{hyperref}
\usepackage{amssymb, amsmath}
\usepackage{graphicx, caption}
\usepackage{booktabs}

\usepackage{natbib}

\usepackage{pdfpages}

\begin{document}
\maketitle

\begin{abstract}
Basket trials have emerged as a new class of efficient approaches in oncology to evaluate a new treatment in several patient subgroups simultaneously. In this paper, we extend the key ideas to disease areas outside of oncology, developing a robust Bayesian methodology for randomised, placebo-controlled basket trials with a continuous endpoint to enable borrowing of information across subtrials with similar treatment effects. After adjusting for covariates, information from a complementary subtrial can be represented into a commensurate prior for the parameter that underpins the subtrial under consideration. 
We propose using distributional discrepancy to characterise the commensurability between subtrials for appropriate borrowing of information through a spike-and-slab prior, which is placed on the prior precision factor.
When the basket trial has at least three subtrials, commensurate priors for point-to-point borrowing are combined into a marginal predictive prior, according to the weights transformed from the pairwise discrepancy measures. In this way, only information from subtrial(s) with the most commensurate treatment effect is leveraged. The marginal predictive prior is updated to a robust posterior by the contemporary subtrial data to inform decision making. 
Operating characteristics of the proposed methodology are evaluated through simulations motivated by a real basket trial in chronic diseases. 
The proposed methodology has advantages compared to other selected Bayesian analysis models, for (i) identifying the most commensurate source of information, and (ii) gauging the degree of borrowing from specific subtrials.
Numerical results also suggest that our methodology can improve the precision of estimates and, potentially, the statistical power for hypothesis testing. 
\end{abstract}


\section{Introduction}
\label{sec:Intro}

There has been an increasing interest in precision medicine (\citealp{NEJM:Mirnezami2012, Nature:Schork2015}) over the past few decades. Rapid advances in genomics and biomarkers allow stratification of patients into subgroups that may have different benefit from new treatments. 
Unlike the one-size-fits-all concept in conventional paradigms of clinical drug development, the aim of precision medicine is to target the right treatments to the right patients at the right time.
In the era of precision medicine, new trial designs have been developed, several of which are examples of {\em master protocols} (\citealp{NEJM:Woodcock2017, JBS:Renfro2018}) to study multiple diseases or multiple agents, or sometimes both. One well known class of {\em master protocols} is basket trials  (\citealp{AO:Renfro2017}). In the simplest formulation, basket trials evaluate a single targeted agent to patients that share a common feature, such as similar genetic mutation, but may present various disease subtypes.
It is administratively more efficient to plan a basket trial than a number of separate trials for the small subgroups, respectively. With one subtrial performed in each patient subgroup, basket trials are advantageous also for addressing multiple research questions simultaneously, for example, which subgroup(s) of patients may benefit and to what extent.  To date, sophisticated approaches to the design and analysis of basket trials have predominantly been proposed for early phase oncology drug development, where the `standard' approach is a single-arm design with binary RECIST endpoint \citep{EJC:Eisenhauer2009, EJC:Schwartz2016}. This manuscript extends the key ideas of basket trials to disease areas outside of oncology, for example, in cases where patients have distinct clinical conditions but share similar symptoms. For this, we develop efficient approaches for analysing randomised, placebo-controlled basket trials which collect data on a continuous endpoint. \\

When analysing early phase basket trials, one major concern is the potential heterogeneity of the treatment effect in various patient subgroups.
Investigators are faced with the dilemma of discarding or incorporating data from other subgroups to reach a decisive conclusion about the treatment effect for a specific subgroup.
The option of using complementary data from subtrials that run concurrently is intriguing, as it may lead to considerable increase in the statistical power of the study to detect drug activity in one or more subgroups. This should be balanced with the risk that treatment effect in an important patient subgroup may be overlooked or missed.
Conventional analysis strategies such as {\em stand-alone analyses} (also known as the approach of no borrowing) and {\em complete pooling} irrespective of subgroup labels have been criticised.
Some authors have proposed using hierarchical random-effects models, as a compromise between the two limiting opinions, to enable borrowing of information across subgroups (\citealp{SiM:Thall2003, COO:Thall2008, CT:Berry2013}).
Such well-established approaches for information borrowing are justified, under the assumption of the exchangeability (\citealp{MS:Bernardo1996}) of subgroup-specific treatment effects. More specifically, exchangeability means that the magnitude of clinical benefit may differ, but nothing is known {\em a priori} to suppose patients of some subgroups benefit better than others.
\citet{PS:Neuenschwander2016} discuss a robust extension to the standard hierarchical models by including the possibility of non-exchangeability for each parameter (vector) that underpins a subgroup.
Their approach permits an extreme subgroup not to be overly influenced by other subgroups in situations of data inconsistency. \\

Additional concerns about the subgroup effect are essential in precision medicine.
Often, the targeted therapy is effective only in some subgroups, and certain subgroups illustrate more similar clinical benefit between themselves than with others.
Several variations of standard hierarchical modelling have been considered suitably for the context of basket trials to implement borrowing of information (\citealp{CCT:Liu2017, CT:Chu2018}). Modifications  are motivated mainly by (i) justification about plausible clustering of similar subgroups, and (ii) quantification about the magnitude, to which a subgroup-specific parameter should be shrunk towards the mean effect across subgroups.
Most recently, more sophisticated methods in the framework of Bayesian model averaging (\citealp{JASA:Madigan1994, JRSS:Draper1995}) have been applied to analysing basket trials.
\citet{BIOS:Psioda2019} average over the {\em complete} model space, which is constituted by all models for possible configurations of the subgroups that may demonstrate the same or disparate efficacy. In a model that assumes identical treatment effect among certain subgroups, information is pooled across the corresponding subgroups under the assumption of interpatient exchangeability. The number of models to be included in the complete model space for averaging increases exponentially with the number of subgroups involved in the basket trial.
\citet{SiM:Hobbs2018} enumerate all possible subgroup pairs, wherein the parameters are considered to be either exchangeable or non-exchangeable. Using the product of, rather than the individual, prior probabilities for any two subgroups being exchangeable, their method offers considerable computational efficiency with respect to conventional Bayesian model averaging.  \\

In this paper, we propose methodology motivated by a randomised, placebo-controlled phase II basket trial, which is being undertaken in patients with chronic diseases. Patients who share a common disease symptom that the new treatment can potentially improve, will be stratified into subgroups according to their clinical conditions. Efficacy will be recorded on a continuous endpoint. Adjustment for baseline covariate(s) is desirable to allow for a more precise estimate of treatment effect. We develop a Bayesian methodology for borrowing of information across consistent subgroups based on commensurate priors (\citealp{BIOM:Hobbs2011, BA:Hobbs2012}), which are a type of hierarchical model for robust estimation in circumstances of just a small number of complementary studies.
Using it can facilitate inferences with respect to all possible pairwise borrowing of information between $K$ subgroups in a basket trial, which accounts for the level of data commensurability across subtrials.
More explicitly, given any complementary subtrial data, a commensurate prior can be specified for the treatment effect in the subtrial of contemporary analysis interest. It is basically a normal predictive prior centred at the complementary subtrial data parameter, with a precision factor to capture the commensurability of the parameters that underpin the complementary and contemparory subtrials. \\

We explore placing an empirical spike-and-slab prior (\citealp{JASA:Mitchell1988}) on the precision factor, which determines the degree of point-to-point borrowing.
For overcoming a prior-data conflict, we propose using a distributional discrepancy measure to characterise the commensurability of information between any two subtrials.
It could quantify the probability mass to be placed on the `spike' prior for strong borrowing, and that on the `slab' prior for discounting inconsistent information from a complementary subtrial, respectively. 
This discrepancy measure meanwhile discerns the complementary subtrials (when at least three subgroups are involved) according to their relative commensurability, and can therefore encourage differential borrowing of information to estimate the model parameter specific to a subtrial.
The proposed methodology for basket trials is fundamentally different from the existing approaches to information sharing.
It avoids the limiting assumption about exchangeability for parameters of certain or all subtrials, but features the use of a distributional discrepancy measure to inform the borrowing only from the most commensurate subtrial(s). \\

The remainder of this paper is structured as follows. We describe the motivating example and decision criteria in Section \ref{sec:motiva}.
In Section \ref{sec:methods}, we present our analysis methodology and discuss how a discrepancy measure may help make appropriate use of complementary data in a basket trial.
In Section \ref{sec:sims}, we perform a simulation study to evaluate the operating characteristics of phase II basket trials that would have been analysed using the proposed methodology, and compare our Bayesian model with some alternative analysis models.
We close with a discussion of our findings and future research that arises in Section \ref{sec:discuss}.

\section{Motivating example and notation}
\label{sec:motiva}

We use a randomised, placebo-controlled phase II basket trial, as a motivating example, which evaluates a new treatment for cognitive dysfunction in patients of primary biliary cholangitis (PBC) and Parkinson's disease (PD).
This clinical trial is led by Newcastle University; at the time of writing, it has been funded but not yet opened to participants.
By stages and types of the chronic diseases, patients are to be recruited and stratified into three disjoint subgroups, that is, early-stage PBC, late-stage PBC, and PD. 
The PBC and PD basket trial thus comprises three subtrials. 
A continuous outcome measuring cognitive performance will be used as the clinical endpoint in each subtrial.
Once the trial begins, patients within each subtrial will be randomised to receive either the new treatment or a placebo. \\

For each subtrial $k=1, \dots, K$, we suppose $n_k$ patients are to be recruited. 
The binary indicator with respect to the treatment assignment is denoted by $T_{ik}$ for patient $i = 1, \dots, n_k$. Specifically, $T_{ik}=1$ if patient $i$ in subtrial $k$ is allocated to the treatment and 0 if a placebo. From each patient, data are measured on covariates, denoted by $\boldsymbol{z}_{ik} = \{z_{ik1}, \dots, z_{ikq} \}$, and a post-treatment outcome of clinical interest, denoted by $y_{ik}$. We fit a linear regression model:
\begin{equation}
\mathbb{E}(y_{ik} |\boldsymbol{z}_{ik}, T_{ik}) = \boldsymbol{z}_{ik}'\gamma_k + T_{ik}\theta_k,
\label{eq:linear}
\end{equation}

\noindent where $\gamma_k$ is a $(1\times q)$ coefficient vector representing the main effects of the covariates, and $\theta_k$ for the treatment effect in a subtrial $k$ is of our primary interest. Within each subtrial
\begin{equation*}
\begin{split}
\mathbb{E}(y_{ik} | \boldsymbol{z}_{ik}, T_{ik}=1) &= \boldsymbol{z}_{ik}'\gamma_k + \theta_k,  \\
\mathbb{E}(y_{ik} | \boldsymbol{z}_{ik}, T_{ik}=0) &= \boldsymbol{z}_{ik}'\gamma_k 
\end{split}
\end{equation*}

\noindent leads to an estimator of the treatment effect over a placebo, denoted by $\Delta (T_{ik}) = \theta_k$.  \\

%The principal aim of this basket trial is to estimate the treatment effect per module. 
A more accurate estimate for $\theta_k$ helps to support the decision as to whether a phase III trial of the treatment should go ahead, and in which patient subpopulation(s).
Moreover, inferences based on evidence of the basket trial can inform the design of a future trial, such as computing the sample size to sufficiently power the trial. 
With respect to either continuing or halting the clinical evaluation, trial decisions per subtrial can be framed as a decision between {\em Go} and {\em No-go}. 
Given a hypothesis for the treatment effect per subtrial $k$:
\begin{equation*}
H_0: \theta_{k} \leq 0 \qquad \text{versus}  \qquad H_a: \theta_{k} > 0,
\end{equation*}

\noindent investigators may base the trial decision on probabilistic inferences about $\theta_k$ to a threshold $\delta_U$, which represents the magnitude of improvement required to declare a clinical benefit of the new treatment, in the Bayesian framework (\citealp{JRSSA:Spiegelhalter1994}).
One example is to compute the posterior probability of $\theta_k$ exceeding the threshold $\delta_U$, in which it is not uncommon to pre-specify $\delta_U$ as a value greater than 0.
Here, we define a formal decision criterion: (i) a {\em Go} decision will be taken for subtrial $k$ if $\mathbb{P}(\theta_k > \delta_U) > \zeta$, otherwise (ii) a {\em No-go}.
This criterion refers to a quantity $\zeta \in [0, 1]$ as the level of evidence that would be required for the new treatment to compellingly provide an improvement over the control.
Choices such as $\zeta = 0.90$ may be appropriate. 
In what follows, we develop a novel Bayesian model that leverages complementary data from the most consistent subtrials for estimating $\theta_k$.

\section{Methods}
\label{sec:methods}

Suppose the patient-level trial data can be modelled using a linear regression model in the form of \eqref{eq:linear}, and $\theta_k$ specific to a subtrial is a continuous location parameter. 
Letting $\boldsymbol{x}_k$ denote the data from subtrial $k$ and $\pi_{0k}(\theta_k)$ the initial vague (operational) prior, the information that subtrial $k$ carries can be represented by an operational posterior,
\begin{equation}
\pi_k (\theta_k | \boldsymbol{x}_k) \propto \mathcal{L}(\boldsymbol{x}_k | \theta_k) \pi_{0k}(\theta_k).
\label{eq:oppost}
\end{equation}

\noindent We label the subtrial of our contemporary analysis interest with $k^\star$.
For estimating each $\theta_{k^\star}$, information from a complementary subtrial $k\neq k^\star$ can be leveraged through a commensurate predictive prior (CPP). Inspired by \citet{BIOM:Hobbs2011, BA:Hobbs2012}, 
we introduce a precision parameter, denoted by $\nu_{kk^\star}$, which parameterises the consistency between $\theta_k$ and $\theta_{k^\star}$.
Conditional on the information commensurability, each predictive distribution for $\theta_{k^\star}$ can be stipulated as
\begin{equation}
\theta_{k^\star} | \theta_k, \nu_{kk^\star} \sim N(\theta_k, 1/\nu_{kk^\star}^2),
\end{equation}
\noindent where $\theta_k$ is inferred from the operational posterior in \eqref{eq:oppost} and the unknown $\nu_{kk^\star}$ determines the degree of borrowing.
This then leads to a CPP as
\begin{equation}
\pi^\text{CPP}(\theta_{k^\star}, \nu_{kk^\star} | \boldsymbol{x}_k, \theta_k) \propto \mathcal{L}(\boldsymbol{x}_k | \theta_k) \pi_{0k}(\theta_{k}) \times \nu_{kk^\star}\phi((\theta_{k^\star} - \theta_k)\nu_{kk^\star})\times g_k(\nu_{kk^\star}),
\label{eq:CPP}
\end{equation}

\noindent where $\phi(\cdot)$ is the standard normal probability density function, and $g_k(\nu_{kk^\star})$ is the prior for $\nu_{kk^\star}$.
If the complementary subtrial $k$ is consistent with $\nu_{kk^\star}\gg 0$, the marginal CPP for $\theta_{k^\star}$ converges to the operational posterior $\pi_k(\theta_k | \boldsymbol{x}_k)$ that is updated from $\pi_{0k}(\theta_k)$, so that the complementary data $\boldsymbol{x}_k$ will be largely incorporated into subtrial $k^\star$. Otherwise, with $\nu_{kk^\star}\approx 0$, $\boldsymbol{x}_k$ are discarded and the marginal CPP for $\theta_{k^\star}$ tends towards the operational prior for $\theta_{k^\star}$. \\

A spike-and-slab distribution (\citealp{JASA:Mitchell1988}) has been found suitable as a prior for the normal precision parameter (\citealp{BA:Hobbs2012, CT:Hobbs2013}). This is a discrete mixture prior with two components, which can provide us a means for robust borrowing.
Specifically, we define $g_k(\nu_{kk^\star})$ as locally uniform between two limits, $0\leq \mathcal{B}_1 < \mathcal{B}_2$, except some portion of probability mass concentrated at a point, $\mathcal{S} > \mathcal{B}_2$, such that
\begin{equation}
\begin{split}
\mathbb{P}(\nu_{kk^\star} < \mathcal{B}_1) &= 0 \\
\mathbb{P}(\nu_{kk^\star} < u) &= w_{kk^\star} \cdot \frac{u - \mathcal{B}_1}{\mathcal{B}_2 - \mathcal{B}_1},  \qquad \mathcal{B}_1 \leq u \leq \mathcal{B}_2, \\
\text{and  } \mathbb{P}(\nu_{kk^\star} > \mathcal{B}_2) &= \mathbb{P}(\nu_{kk^\star} = \mathcal{S}) = 1 - w_{kk^\star},
\end{split}
\label{eq:spikeslab}
\end{equation}

\noindent where $w_{kk^\star}$ is the probability that $\nu_{kk^\star} \sim \text{Unif}(\mathcal{B}_1, \mathcal{B}_2)$.
Given a subtrial $k$ with sufficiently consistent treatment effect, we expect strong borrowing from $\boldsymbol{x}_k$.
This requires that the normal precision parameter $\nu_{kk^\star}$ in \eqref{eq:CPP} takes a large value, which is possible when the `spike' prior has a large probability mass, that is, when $w_{kk^\star}$ is sufficiently small. Otherwise, $\boldsymbol{x}_k$ is down-weighted by allocating more probability mass to the `slab' prior. We interpret $w_{kk^\star}$ as our prior opinion (on the probability scale) about the pairwise subtrial incommensurability.\\

To determine $w_{kk^\star}$, we propose using a discrepancy measure that quantifies the distributional divergence between the posterior density distributions, $\pi_k(\theta_k | \boldsymbol{x}_k)$ and $\pi_{k^\star}(\theta_{k^\star} | \boldsymbol{x}_{k^\star})$, arising from the same operational prior. 
One viable option is the Hellinger distance (\citealp{Hellinger:Dey1994}):
\begin{equation}
d_{H}(\pi_{\theta_k}, \pi_{\theta_{k^\star}}) = \sqrt{\frac{1}{2}\int_{-\infty}^\infty \left(\sqrt{\frac{{\rm d} \pi_{k^\star}(\theta_{k^\star} | \boldsymbol{x}_{k^\star})}{{\rm d} \theta}}-\sqrt{\frac{{\rm d} \pi_k(\theta_k | \boldsymbol{x}_k)}{{\rm d} \theta}}\right)^2 {\rm d} \theta}.
\end{equation}
\noindent Derivable from the Cauchy-Schwarz inequality, the computed Hellinger distance $d_{\phi_H}(\pi_{\theta_k}, \pi_{\theta_{k^\star}})$ will strictly fall into the interval of [0, 1], which is convenient for characterising the probability that treatment effects of any two subtrials are regarded as dissimilar.
We may then relate the `slab' prior probability $w_{kk^\star}$ with the computed Hellinger distance, simply by stipulating $w_{kk^\star} = d_{H}(\pi_{\theta_k}, \pi_{\theta_{k^\star}})$. In an extreme case that two subtrials are perfectly consistent, i.e., $d_{H}(\pi_{\theta_k}, \pi_{\theta_{k^\star}}) \to 0$, the whole probability mass will be concentrated at the `spike', $\mathcal{S}$.
In turn, this will result in a notably small normal variance $1/\nu_{kk^\star}^2$ of the CPP in \eqref{eq:CPP} such that the complementary subtrial data $\boldsymbol{x}_k$ can be fully incorporated. 
In addition, knowing the upper and lower bounds, which are 0 and 1 for the Hellinger distance, makes it easier to standardise a collection of pairwise discrepancy measurements. This can help quantify the relative importance of all other subtrials $k \neq k^\star$ to form a prior for $\theta_{k^\star}$ in circumstances of $K\geq 3$.
More importantly, the Hellinger distance is preferred over other distributional discrepancy measures, because of its desirable properties of symmetry and invariance to any transformation, for example, logarithmic, exponential, or inverse of square root, of both densities (\citealp{Jeffreys1961}). 
As a symmetric measure of discrepancy, $d_{H}(\pi_{\theta_k}, \pi_{\theta_{k^\star}}) = d_{H}(\pi_{\theta_k^\star}, \pi_{\theta_{k}})$. In a basket trial with $K=2$ and no {\em a priori} assumption about which subtrial has stronger treatment effect, using the Hellinger distance to define the spike-and-slab prior will result in the same magnitude of down-weighting or leveraging subtrial data $\boldsymbol{x}_1$ to subtrial 2 and $\boldsymbol{x}_2$ to subtrial 1. Whilst the invariance property ensures that the computed Hellinger distance $d_{H}(\pi_{\theta_k}, \pi_{\theta_{k^\star}})$ truly reflects the discrepancy between the treatment effect distributions in different patient subtrials, when the linear regression model \eqref{eq:linear} may be parameterised in a different way, for example, with the treatment effect represented by the exponential of $\theta_k$. Given the invariance, we know that
$d_{H}(\pi_{\theta_k}, \pi_{\theta_{k^\star}}) = d_{H}(\pi_{\exp(\theta_k)}, \pi_{\exp(\theta_{k^\star})})$. \\

For any $\theta_{k^\star}$, there exist $(K-1)$ complementary subtrials as the sources where the possible values can be drawn upon.
We now turn our attention to combining the $(K-1)$ CPPs for synthesising information from these complementary subtrials for basket trials with $K\geq 3$. 
The CPPs for point-to-point borrowing are robust in that inconsistent information from any subtrial $k \neq k^\star$ can be down-weighted through the `slab' prior placed on each $\nu_{kk^\star}$, incorporating the use of pairwise Hellinger distance.
If values of $\nu_{kk^\star}$ are appropriately specified, we may obtain $(K-1)$ normal predictive priors marginally on $\theta_{k^\star}$:
\begin{equation}
\int \pi^\text{CPP}(\theta_{k^\star}, \nu_{kk^\star} | \boldsymbol{x}_k, \theta_k) {\rm d}\nu_{kk^\star} \propto \int \mathcal{L}(\boldsymbol{x}_k | \theta_k)\pi_{0k}(\theta_k)\times\nu_{kk^\star}\phi ((\theta_{k^\star} - \theta_k)\nu_{kk^\star})g_k(\nu_{kk^\star}) {\rm d}\nu_{kk^\star},
\end{equation}
\noindent where each one separately may be represented as a $N(\lambda_k, \xi_k^2)$ distribution for the ease of notations.
We further see $\theta_{k^\star}$ as a weighted sum of $(K-1)$ hypothetical random variables:
\begin{equation}
\theta_{k^\star} = \sum_{k=1,\, k\neq k^\star}^K p_{kk^\star} \tilde{\theta}_k, \quad \text{ for } k^\star = 1, \dots, K,
\label{eq:sumtheta}
\end{equation}

\noindent where we suppose each $\tilde{\theta}_k \sim N(\lambda_k, \xi_k^2)$ and the weight vector $\boldsymbol{p}_{k^\star} = (p_{1k^\star}, p_{2k^\star}, \dots)$, containing $(K-1)$ elements, satisfies $\sum \boldsymbol{p}_{k^\star} = 1$.
This further gives 
\begin{equation}
\theta_{k^\star} | \boldsymbol{x}_{(-k^\star)} \sim N \left( \sum_{k=1,\, k\neq k^\star}^K p_{kk^\star} \lambda_k, \sum_{k=1,\, k\neq k^\star}^K p_{kk^\star}^2 \xi_k^2 \right),
\label{eq:combCPP}
\end{equation}
\noindent in which $\boldsymbol{x}_{(-k^\star)}$ denotes the entire trial data excluding those from subtrial $k^\star$. This allows for leveraging information from multiple sources. 
This marginal predictive prior, denoted by $\pi^\text{MPP}(\theta_{k^\star} | \boldsymbol{x}_{(-k^\star)})$, is updated to the posterior with the subtrial data $\boldsymbol{x}_{k^\star}$ using Bayes' Theorem:
\begin{equation}
\pi^\text{MPP}(\theta_{k^\star} | \boldsymbol{x}_{k^\star}, \boldsymbol{x}_{(-k^\star)}) \propto \mathcal{L}(\boldsymbol{x}_{k^\star} | \theta_{k^\star})\times \pi^\text{MPP}(\theta_{k^\star} | \boldsymbol{x}_{(-k^\star)}),
\end{equation}
\noindent so as to inform decision making for subtrial $k^\star$. \\

To allocate a sensible weight to each complementary subtrial, we expect $p_{kk^\star}$ to take large values (close to 1) if subtrials $k$ and $k^\star$ are consistent. 
Specification of these weights $p_{kk^\star}$ may be guided by the  Hellinger distance, $d_{H}(\pi_{\theta_k}, \pi_{\theta_{k^\star}})$ (labelled as $d_{kk^\star}$ for notation simplicity), which we use to measure the commensurability between subtrials $k$ and $k^\star$.
For a basket trial with $K$ subtrials, the Hellinger distances can be organised as a $K\times K$ symmetric matrix:
$$
\begin{pmatrix} 
0 & d_{12} &\cdots & d_{1K} \\
d_{21}  & 0  &\cdots  & d_{2K} \\
\vdots  & \vdots  & \ddots & \vdots \\
d_{K1}  & d_{K2}  &\cdots  & 0
\end{pmatrix},
$$
where each column $k^\star=1,\dots,K$ describes the pairwise distributional discrepancy between our target parameter $\theta_{k^\star}$ and a source parameter $\theta_k$, for $k \neq k^\star$.
Hellinger distances contained in each column $k^\star$ can be normalised into a series of weights $p_{kk^\star} \in [0, 1]$. For this, we will simply stipulate a decreasing function of $d_{kk^\star}$ as
\begin{equation}
p_{kk^\star} = \frac{\exp(-d_{kk^\star}/s_0)}{\sum_k \exp(-d_{kk^\star}/s_0)},
\label{eq:probHD}
\end{equation}

\noindent where a pre-defined $s_0$ governs how much influence the Hellinger distance has on the weight to be computed. 
With a value of $s_0\gg d_{kk^\star}$, nearly the same weight will be allocated irrespective of the pairwise Hellinger distances that could be very different from each other. Whereas, with $s_0 \to 0^+$, the weight corresponding to a Hellinger distance close to 0 tends to be 1.
In Section A of the Web-based Supplementary Materials, we illustrate properties of this transformation analytically in more details.
Weights converted from the pairwise Hellinger distances following \eqref{eq:probHD} can then be assigned to each $\tilde{\theta}_k$, as was stipulated in \eqref{eq:sumtheta}. \\

We would like to add one more note here. The stipulation of weights $p_{kk^\star}$ summing to 1 does not restrict the potential of full borrowing of information in situations, where all the $(K-1)$ complementary subtrials are perfectly consistent with subtrial $k^\star$.
In such a scenario, the Hellinger distance $d_{kk^\star} = 0$, suggesting that 
the CPPs marginally on $\theta_{k^\star}$, represented by $N(\lambda_k, \xi_k^2)$, have identical mean and variance to those of $\pi_{k^\star} (\theta_{k^\star} | \boldsymbol{x}_{k^\star})$. Moreover, equal weights, i.e., $p_{kk^\star} = \frac{1}{K-1}$, will be allocated to the complementary subtrials, respectively. Following \eqref{eq:combCPP}, a predictive prior $\pi^\text{MPP}(\theta_{k^\star} | \boldsymbol{x}_{(-k^\star)})$ would be obtained with its mean as $\lambda_k$ and variance as $\frac{1}{K-1} \xi_k^2$.
With the inclusion of $\boldsymbol{x}_{k^\star}$, the posterior mean and variance become $\lambda_k$ and $\frac{1}{K} \xi_k^2$, respectively.
This indicates all the complementary subtrial data $\boldsymbol{x}_{(-k^\star)}$ have been fully incorporated, and our methodology converges to the approach of {\em complete pooling} in the case of perfect information consistency. 

\section{Simulation study}
\label{sec:sims}

In this section, we illustrate applications of the proposed analysis methodology, and compare it with alternative Bayesian models that may be used for analysing basket trials through a simulation study. Our trial examples are hypothetical, but can represent the situation of a phase II basket trial, for which the analyses are performed to enable sharing of information. 
The main characteristics of the study we simulate are based on the motivating PBC and PD trial described in Section \ref{sec:motiva}. For illustrative purposes, we assume six subtrials instead of three, as typically a fairly large number of patient subgroups would be examined; see, for example, \citet{NEJ:Hyman2015, Nature:Hyman2018} report the results from basket trials with six and nine subtrials, respectively.

\subsection{Basic settings}

We simulate basket trials with $K=6$ subtrials of unequal sample sizes: $n_k \in \{10, 10, 14, 16, 20, 20 \}$, respectively. 
Treatment allocation of individual patients follows a block approach by subtrial, with exactly half of the sample size of each subtrial to receive the treatment and placebo.
We simulate two covariates for each patient as $z_{ik1} \sim N(6, 0.2^2)$ and $z_{ik2} \sim N(4, 0.2^2)$, for $i = 1, \dots, 90$. In particular, $z_{ik1}$ is assumed to be the baseline measurement of the clinical endpoint at the time of randomisation. 
We generate the trial data from a linear regression: for $i = 1, \dots, 90, \, k = 1, \dots, 6$, 
\begin{equation}
\begin{split}
y_{ik} &\sim N(\eta_{ik}, \sigma^2) \quad \text{i.i.d.} \\
\eta_{ik} &= \gamma_{0k} + z_{ik1} \gamma_{1k} + z_{ik2} \gamma_{2k} + T_{ik}\theta_k,
\end{split}
\end{equation}

\noindent where we set the `true' parameter values for the intercept and effect of baseline covariates to $\gamma_{0k} = 5$, $\gamma_{1k} = 3$ and $\gamma_{2k} = 1.3$, and the inter-patient standard deviation $\sigma = 0.4$ to generate the data.
Basket trials are simulated under nine scenarios (listed in Table \ref{tab:scenarios}), which feature varying treatment effect sizes $\theta_k$ and different degrees of heterogeneity across subtrials. All sets of the `true' values of $\theta_k$ are realisations from distinct multivariate normal distributions: we stipulated a high pairwise correlation coefficient (0.8) for $\theta_k$ of the consistent subtrials and a low pairwise correlation coefficient (0.1) between $\theta_k$ of an extreme subtrial and of one else from the rest.
In particular, scenarios 7 and 8 can be seen as `mixed null' scenarios, and scenario 9 is a `global null'. \\

When fitting the Bayesian analysis models, we consider random effects for $\gamma_{0k}$, $\gamma_{1k}$ and $\gamma_{2k}$:
$$
\gamma_{0k} = \chi_0 + \epsilon_0^2,  \quad \gamma_{1k} = \chi_1 + \epsilon_1^2  \quad \text{ and } \quad \gamma_{2k} = \chi_2 + \epsilon_2^2,
$$
setting an uninformative normal prior $N(0, 5^2)$ on each $\chi_j$ and a half-normal prior $HN(z)$ on each $\epsilon_{j}$, for $j = 0, 1, 2$. Here, $HN(z)$ is defined as a $N(0, z^2)$, truncated to cover the interval $(0, \infty)$. The use of a half-normal prior is consistent with the recommendation by \citet{CT:Cunanan2019}. We stipulate $\epsilon_j \in HN(1)$, of which the prior and 95\% credible interval are 0.674 and (0.031, 2.241), to permit very limited information borrowing across subtrials for estimating $\gamma_{jk}, \, j=0, 1, 2,\, k = 1, \dots, 6$. In the following, we describe additional specifications to implement the Bayesian models that estimate $\theta_k$ with or without information leveraged from other subtrials.\\

To implement our methodology for estimating $\theta_k$, we choose setting $\mathcal{B}_1 = 0.01, \, \mathcal{B}_2 = 1$ and $\mathcal{S} = 100$ for the spike-and-slab prior on each $\nu_{kk^\star}$. The `slab' prior is very uninformative and is sufficient to fully discard the entire information from an external subtrial $k$; the `spike' prior is specified so that the proposed methodology can be reduced to {\em complete pooling} in situations of perfect information commensurability.
Justification of choosing this spike-and-slab prior is provided in Section B of the Web-based Supplementary Materials.
An initial vague prior $\pi_{0k}(\theta_k)$ is used for $\theta_k, \, k = 1, \dots, 6$; we use $N(0, 10^2)$ such that the 95\% prior credible interval is (-19.560, 19.560), covering a wide range of possible $\theta_k$. 
To yield a large (small) weight $p_{kk^\star}$ corresponding to a small (large) Hellinger distance, we let $s_0 = 0.15$ to leverage information from all other subtrials. Nevertheless, we study how different stipulations of $s_0$ may impact on the identification of the most consistent subtrial(s) in Section C of the Supplementary Materials, exploring $s_0 = 0.25, 0.35, 0.45$ in addition. We are interested in comparing the proposed methodology with 
\begin{enumerate}
\item Standard hierarchical model (HM) that assumes fully exchangeable parameters: $\theta_k | \mu_, \tau \sim N(\mu, \tau^2)$ with $\mu \sim N(0, 10^2)$ and $\tau\sim HN(0.125)$. The median and 95\% credible interval of $HN$(0.125) are 0.084 and (0.004, 0.280), respectively. 
\item Bayesian model with no borrowing of information. Trial data are stratified by subtrials for stand-alone analyses, setting each $\theta_k \sim N(0, 10^2)$. Random effects for $\gamma_{0k}$, $\gamma_{1k}$ and $\gamma_{2k}$ therefore cannot be estimated; we then place a $N(0, 5^2)$ prior on each.
\item EXNEX model by \citet{PS:Neuenschwander2016}, with equal prior probabilities of exchangeability (EX) and non-exchangeability (NEX). The EX distribution has the same parameter configuration as what was stipulated for the standard HM above, and the six NEX distributions are all set to be $N(0, 10^2)$.
\end{enumerate}

Comparison is in terms of the precision of their posterior point estimates, more specifically, the posterior means, for $\theta_k$ that could be measured by an analogue of bias and mean squared error (MSE):
\begin{equation*}
\begin{split}
\text{Bias}(\theta_k) &\approx \frac{1}{M} \sum_{m = 1}^M \bar{\theta}_k^m - \theta_k, \\
\text{MSE}(\theta_k) &\approx \frac{1}{M} \sum_{m = 1}^M (\bar{\theta}_k^m - \theta_k)^2,
\end{split}
\end{equation*}
\noindent where $M$ is the total number of replicates in the simulation study, and $\bar{\theta}_k^m$ denote the posterior means of $\theta_k$ for the $m$-th simulated basket trial. These metrics will be reported by subtrial.
We also compare these Bayesian analysis models with respect to the trial operating characteristics, such as the subtrial-wise error rates.
Corresponding to the frequentist type I error rate and statistical power, we will report proportions of the simulated trials with
\begin{itemize}
\item an erroneous {\em Go} decision in a subtrial for the `true' $\theta_k =0$, and
\item a correct {\em Go} decision in a subtrial for the `true' $\theta_k > 0$,
\end{itemize}

\noindent respectively. An overall (analogue of) type I error rate, defined as erroneous {\em Go} decision made for at least one subtrial, will also be reported for scenarios with null $\theta_k$'s. In particular, a {\em Go} will be allocated if $\mathbb{P}(\theta_k > \delta_U) > 0.975$. Trial operating characteristics will be evaluated setting $\delta_U=0.25, 0.30$. We are especially interested in Scenarios 7 -- 9, the mixed or global null scenarios, to report the (analogue of) type I error rate by both subtrial and overall. \\

Results will be summarised by averaging across 10,000 replicates of the basket trial. 
The Bayesian analysis models are fitted in R version 3.4.4 using the R2OpenBUGS package based on two parallel chains, with each running the Gibbs sampler for 10,000 iterations that follow a burn-in of 3000 iterations.
OpenBUGS code, together with R functions, to implement each of the Bayesian analysis models is available at \url{https://github.com/BasketTrials/Bayesian-analysis-models}.

\subsection{Results}

\begin{figure}[!p]
\centering
%\captionsetup{font=scriptsize}
\includegraphics[width=1\linewidth]{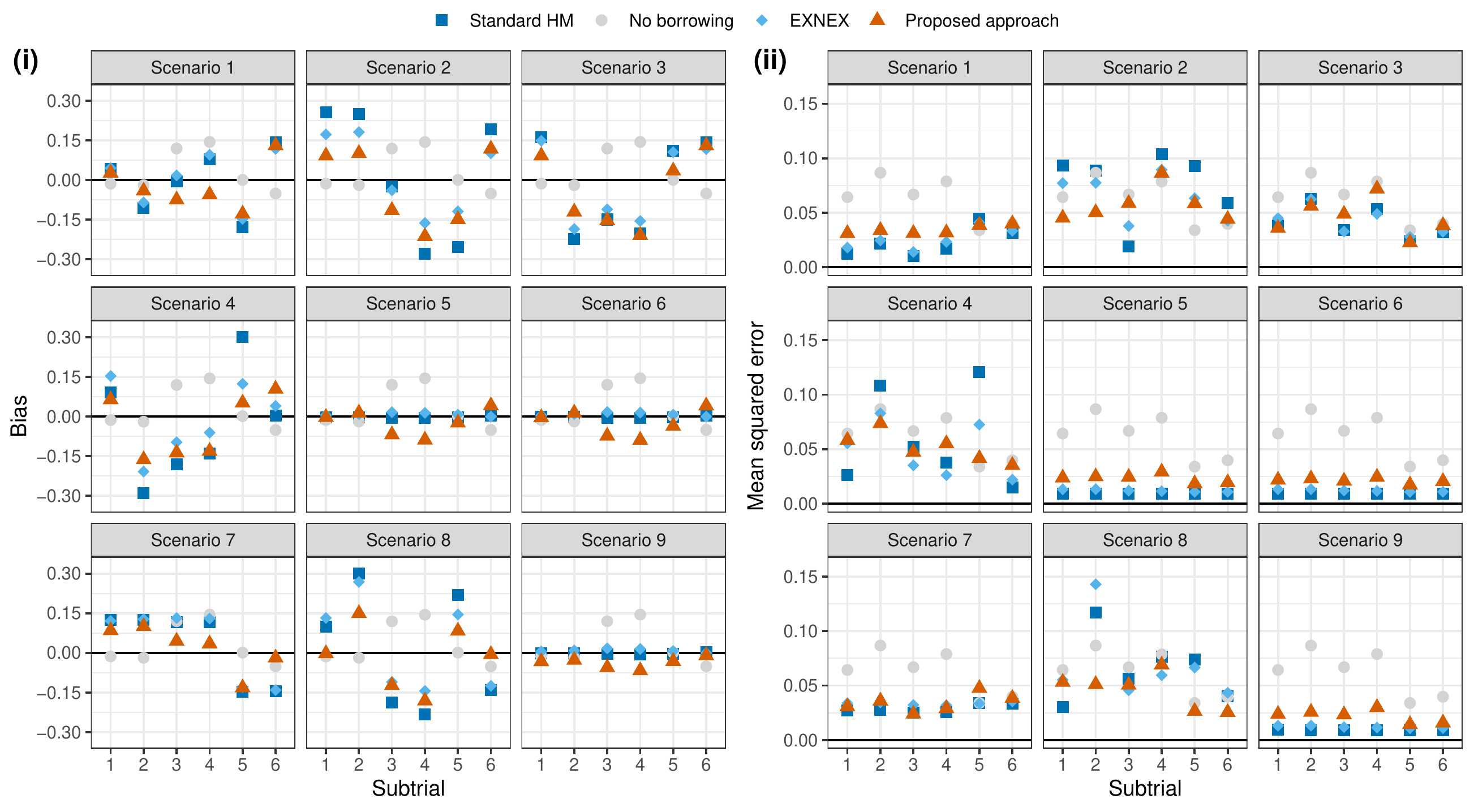}
\caption{Bias and mean squared error of the posterior estimators for $\theta_k$ based on the Bayesian models.}
\label{fig:BiasMSE}
\end{figure}

Figure \ref{fig:BiasMSE} compares the performance of the posterior estimators yielded by the Bayesian models.
It shows that the proposed methodology produces smaller bias and MSE than the standard HM and EXNEX, across nearly all scenarios. 
Point estimators based on the standard HM and EXNEX work well in scenarios 5, 6 and 9 as the small-to-moderate variability between $\theta_k$s can be addressed by setting $\tau \sim HN(0.125)$.
The proposed analysis methodology, in contrast, distinguishes the heterogeneity more sensitively. 
Much smaller bias and MSE are yielded when estimating $\theta_k$ for basket trials with divergent treatment effects across subtrials; see, for example, scenario 2.
In situations where information from other subtrials should be largely discounted, referring to scenarios 7 and 8, our methodology generates comparatively similar bias to the no borrowing approach but with a smaller MSE.
This is because information from subtrials with a non-zero treatment effect, for example, in scenario 8, can be largely discounted to formulate the marginal predictive prior for $\theta_1$. \\

We have also compared the Bayesian analysis models in terms of the average width of the posterior credible intervals for $\theta_k$. In contrast to the alternative Bayesian models, the proposed methodology yields posterior estimates with narrower credible intervals when there is at least one consistent complementary subtrial; see Figure S1 of the Supplementary Materials. 
When using the proposed analysis methodology for borrowing of information, investigators may be interested in the weight eventually allocated to each external subtrial for obtaining the marginal predictive prior. In Figures S3 and Figure S4 of the Supplementary Materials, we comment with regards to scenarios 4 (divergent $\theta_k$) and 5 (consistent $\theta_k$) on the weight allocation based on the assessed pairwise commensurability, and illustrate how the pre-specified value of $s_0$ impact the sensitivity of the proposed methodology to identify the most commensurate subtrial(s). \\

Table \ref{tab:typeI} quantifies the impact of using different Bayesian models on the error rate control under the null hypothesis.
Here, we report the (analogue of) type I error rate for scenarios involving at least one subtrial with $\theta_k =0$, setting $\delta_U=0.25$. Comparisons where setting $\delta_U=0.30$ are given in Table S1 of the Supplementary Materials.
For scenario 9 (global null), all the four Bayesian analysis models control the error rate well following the decision criterion. 
Nevertheless, the approaches that enable borrowing of information, i.e., standard HM, EXNEX and the proposed methodology, have resulted in smaller type I error rates, compared with the approach of no borrowing, since incorporating consistent information from other subtrials reassures that making a {\em Go} decision is not justified.
Our approach produces slightly higher error rates than standard HM and EXNEX, as for some simulated trials information from subtrials with a similar low treatment effect may be shared (but not with those of a null $\theta_k$'s), leading to a higher chance to reject the null hypothesis.
In scenario 8 where some subtrials have large treatment  effects, we observe a higher increase in the error rate, using standard HM and EXNEX approaches, compared with the proposed approach.
We note that a difference in the sample sizes of subtrials 2 and 4 or 5 (for all scenarios) leads to disparate impact on the error rate of the same approach in the same scenario with null $\theta_k$'s: those for subtrial 2 are regularly larger than subtrial 4 or 5. More explicitly, when reacting to a prior-data conflict, a larger sample size of subtrial 4 or 5 provides more evidence to evaluate the plausibility of down-weighting; estimation of $\theta_4$ or $\theta_5$ thus has increased chances to avoid being overwhelmed by the complementary information. \\

\begin{figure}[!p]
\centering
%\captionsetup{font=scriptsize}
\includegraphics[width=0.75\linewidth]{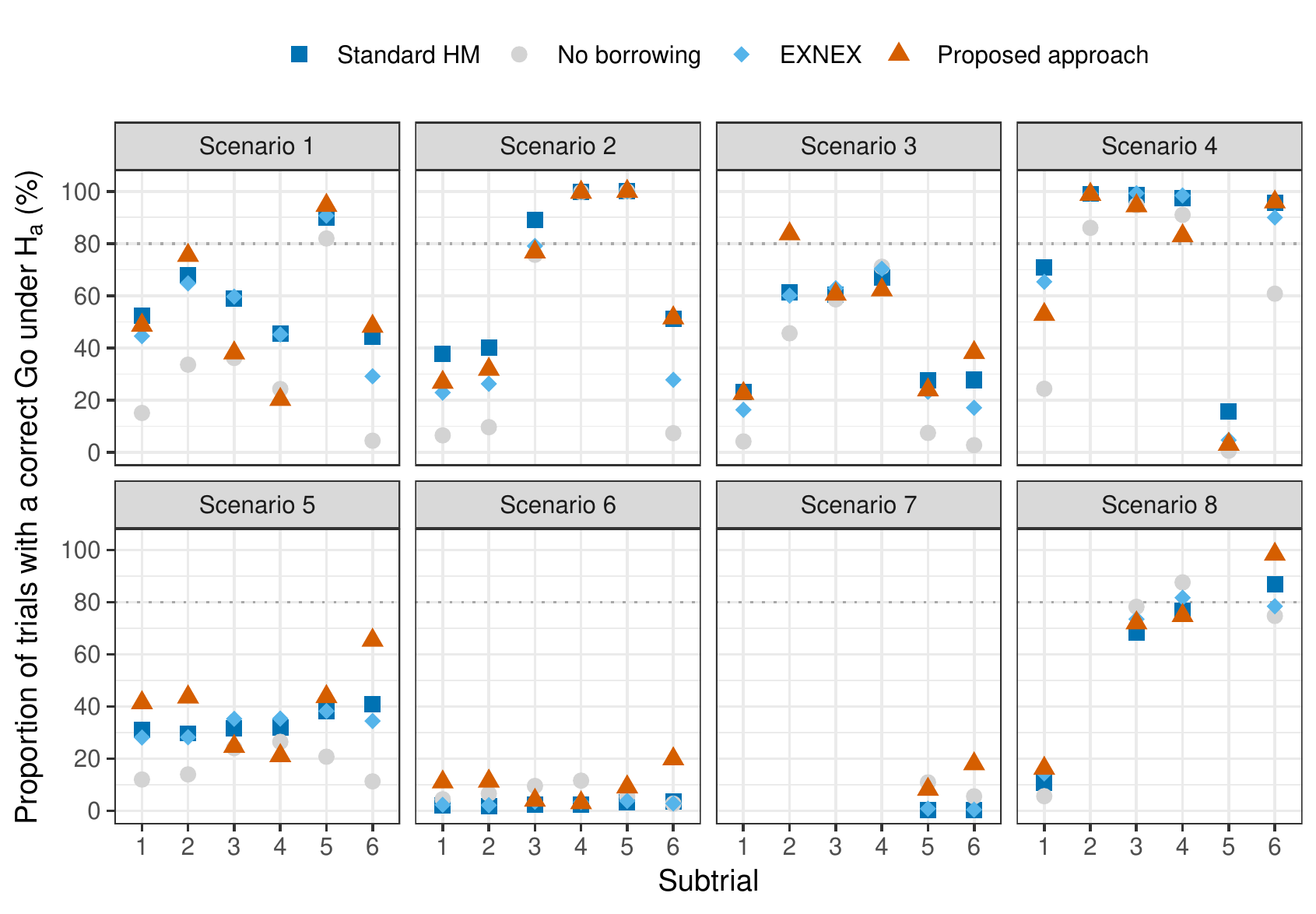}
\caption{Comparison of the Bayesian analysis models with respect to the analogue of statistical power: null hypothesis is correctly rejected in the presence of a treatment effect per subtrial, setting $\delta_U=0.25$ and $\zeta = 0.975$.}
\label{fig:power}
\end{figure}

What may also be interesting to investigators is the potential increase in statistical power to demonstrate the treatment effect, by incorporating information from complementary subtrials.
Figure \ref{fig:power} visualises the comparison of the Bayesian analysis models in terms of correctly declaring a clinical benefit in subtrial $k$, setting $\delta_U=0.25$; see Figure S5 of Supplementary Materials that visualises the comparison setting $\delta_U=0.30$. 
Across nearly all subtrials of the simulated basket trials in scenarios 1 -- 5, the Bayesian approaches of borrowing show substantial advantages over the approach of no borrowing. 
When comparing between the Bayesian approaches of borrowing, we check how the chance would be for a subtrial with comparatively low treatment effect to be concluded with a correct {\em Go} decision, in the presence of consistent subtrials. Looking at scenarios 3, for example, our approach leads to higher statistical power for subtrials 2 and 6 compared with other Bayesian models. \\

Scenarios 5 and 6 represent situations where all subtrials are commensurate, but the former has a larger treatment effect size.
Given our criterion that $\mathbb{P}(\theta_k > 0.25) > 0.975$ for a {\em Go} decision, scenario 6 with all $\theta_k = 0.30$ is particularly a hard scenario to allocate a {\em Go}. 
Compared with the approaches that enable borrowing, using the approach of no borrowing results in subtrials 3 and 4 having a slightly higher probability of correct {\em Go} decision.
However, this does not mean the no borrowing approach is superior, since standard HM and EXNEX produce estimates of $\theta_k$ with a similar level of bias, but much smaller posterior variances than the approach of no borrowing. These results are observed from Figure \ref{fig:BiasMSE} and Figure S2 of the Supplementary Materials. 
More informative posterior distributions for $\theta_k$ nevertheless do not necessarily mean a higher interval probability of $\theta_k>\delta_U$: it is possible that diffuse posteriors for $\theta_k$ obtained from the approach of no borrowing has comparable or even higher chances to exceed the level $\gamma = 0.975$. 
When the consistent `true' $\theta_k$s increase from 0.30 (scenario 6) to 0.45 (scenario 5), we begin to observe the efficiency gains by using Bayesian models that permit borrowing of information than no borrowing. The proposed methodology appears to present a larger absolute gain in power compared with the alternative models, although we note that the absolute gain in power can be a misleading metric due to the non-linear shape of the power curve. \\

In scenario 7, due to the prior specification of $\tau \sim HN(0.125)$ being incapable of accounting for the variability across subtrials, both standard HM and EXNEX shrink $\theta_5$ and $\theta_6$ excessively towards the mean effect across subgroups. This in turn dilutes the treatment effect in corresponding subtrials. Consequently, it seems better to implement the approach of no borrowing for possibility of declaring a positive treatment effect.
Our approach presents slightly higher power than the no borrowing approach as there is some consistent information to be incorporated from a complementary subtrial.
In scenario 8, our approach performs similarly to the alternatives, but slightly better for subtrial 6 due to leveraging consistent information. \\

We note this simulation study does not consider cases of basket trials involving rare disease subgroups, where certain subtrials can have a much smaller sample size than others. We present several hypothetical data examples in Section D of the Supplementary Materials to comment on the sensitivity to the difference in subtrial sample sizes.

\section{Discussion}
\label{sec:discuss}

The paradigm shift towards precision medicine opens new avenues for novel trial designs and analysis methodologies to deliver more tailored healthcare to patients. 
Basket trials emerge as a new class of efficient approaches to oncology drug development in the era of precision medicine, offering a framework to evaluate the treatment effect together with its heterogeneity in various patient subgroups.
In this paper, we have extended the key ideas of a basket trial approach to disease areas outside of oncology, and proposed a new Bayesian model to enable borrowing of information from the most commensurate subtrial(s) without requiring {\em a priori} clustering of similar subgroups.
By including an information discrepancy measure, it can discern the degree of borrowing from complementary subtrials.
In particular, the Hellinger distance plays a dual role in our methodology: (i) it gauges the maximum amount of information that could be leveraged from a specific subtrial $k \neq k^\star$ to estimate $\theta_{k^\star}$; (ii) when there are $K \geq 3$ subtrials, it determines the weight allocation to reflect the relative importance for appropriate borrowing of information. \\

The Bayesian analysis methodology in Section \ref{sec:methods} has been developed assuming the basket trial generates continuous response data. However, it could be easily generalised to analyse other types of data that can be fitted using a generalized linear model for non-Gaussian error distributions. For example, it would be readily applicable to analysing phase II basket trials that use binary endpoints:
after fitting the patient-level data per subtrial with a logistic regression model, our approach may be considered to stipulate commensurate predictive priors, informed by the pairwise Hellinger distance, for the subtrial-specific parameters to discuss borrowing of information from the most consistent subtrial(s).
For down-weighting in cases of data conflict suggested by the Hellinger distance, we did not delve into calibration of the `slab' prior but simply use a very uninformative uniform distribution, which ensures data from an inconsistent subtrial can be discarded.
When using the proposed methodology in practice, we recommend specifying the spike-and-slab prior based on some preliminary knowledge about the magnitude of variances of $\theta_k$.
Specification of the `slab' prior may particularly deserve future research to exploit the advantage of the proposed methodology. We refer to \citet{PS:Mutsvari2016} as a relevant investigation, which focuses on choosing the diffuse component of a mixture prior for robust inferences. The exploration may be closely linked with the users' stipulation of the prior probability weight, which is based upon the Hellinger distance, to be attributed to the `slab' prior. \\

In our simulation study, we have considered imbalance subtrial sizes. Simulation results show that our methodology can down-weight inconsistent information from a subtrial that has larger sample sizes. For illustrative purposes, we have supposed equal randomisation ratio between treatment groups within a subtrial. 
Investigators can pragmatically determine the randomisation ratio  as well as the subtrial-wise sample size for a basket trial that may base decision making on our analysis methodology. 
Potentially, more dosage groups of the same treatment in each subtrial can be considered.
Also, many have shown a great interest in sequential basket trials \citep{SiO:Simon2016, SiM:Cunanan2017, SiM:Hobbs2018} with interim look(s) incorporated for the possibility of, say, terminating enrollment of patients in ineffective subgroups. 
We note that the proposed Bayesian approach can be implemented with any number of analyses following a flexible timescale for interim decision making. 
There is no requirement of a minimum sample size per subtrial to carry out an interim look, due to the use of an initial operational prior $\pi_{0k}(\theta_k)$ for computing the pairwise Hellinger distance. However, an inflation of type I error rate arising from such repeated significance tests would occur. \\

Throughout, we have restricted our focus onto basket trials, where the subtrials use the same endpoint across patient subgroups.
In many disease areas multiple endpoints (\citealp{FDA:FIH2017}) may often arise, as it could involve various dimensions to conclude on the clinical benefit. One common situation is to continue monitoring toxicity in addition to the assessment of efficacy (\citealp{BIOM:Bryant1995, CTT:Tournoux2007}). 
With regards to this, our approach could be extended in several ways.
For instance, in cases where the set of multiple endpoints remain the same across subgroups, it would be straightforward to establish a joint probability model and derive the pairwise Hellinger distance between multivariate probability densities (\citealp{CRC:Pardo2005}). 
Suitable alternatives include separating the discussion about borrowing of information by endpoint. A unified utility function may then be adopted for trial decision making based on evidence on multiple endpoints.
In another more complex setting where the efficacy endpoint, for example, could be distinct but correlated across subgroups, one might need to translate the subtrial data onto a common scale in order to adapt the present approach. Ideas could be drawn from \citet{SMMR:Zheng2019}, where incorporation of supplementary data recorded on a different measurement scale has been discussed in the context of phase I clinical trials.

\bibliographystyle{apalike}
%\bibliography{methods2}

%%%%%%%%%%%%%%%%%%%%%%%%%%%%%%%%%%%%%%%%%%%%%%%%%%

%%%%%%%%%%%%%%%%%%%%%%%%%%%%%%%%%%%%%%%%%%%%%%%%%%

\begin{table}[!p]
\begin{center}
    \caption{Simulation scenarios with specification of the `true' treatment effect $\theta_k$ to compare the Bayesian analysis models. The figure in bold indicates a 0 or low treatment effect.}
    \label{tab:scenarios}
    \begin{tabular}{c|c|c|c|c|c|c} % <-- Alignments: 1st column left, 2nd middle and 3rd right, with vertical lines in between
      & \multicolumn{6}{c}{Subtrial $k$ (Sample size, $n_k$)} \\
      \cline{2-7}
      Scenario & 1 ($n_1 = 10$)  & 2 ($n_3 = 10$)  & 3 ($n_2 = 14$)   & 4 ($n_5 = 16$)  & 5 ($n_4 = 20$)   & 6 ($n_6 = 20$) \\
      \hline
      1  & 0.49 & 0.67  & 0.54  & 0.43 & 0.79   & {\bf 0.35} \\
      2 & {\bf 0.35} & 0.37  & 0.80  & 1.30 & 1.38   & 0.40 \\
      3 & {\bf 0.29}  & 0.77  & 0.68  & 0.75 & {\bf 0.33}  & {\bf 0.30} \\
      4 & 0.59 & 1.17  & 1.02  & 0.95 & {\bf 0.13}   & 0.75 \\
      5 & 0.45 & 0.45  & 0.45  & 0.45  & 0.45 & 0.45 \\
      6 & {\bf 0.30} & {\bf 0.30}  & {\bf 0.30}  & {\bf 0.30}  & {\bf 0.30} & {\bf 0.30} \\
      7 & {\bf 0}    & {\bf 0}     & {\bf 0}    & {\bf 0}    & 0.37   & 0.37 \\
      8 & {\bf 0.33} & {\bf 0}  & 0.82  & 0.90   & {\bf 0}    & 0.83 \\
      {\color{red} 9} & {\bf 0}    & {\bf 0}     & {\bf 0}    & {\bf 0}    & {\bf 0}     & {\bf 0}   \\
      \bottomrule
    \end{tabular}
\end{center}
\end{table}

\begin{table}[!p]
  \begin{center}
    \caption{Comparison of the Bayesian analysis models with respect to the analogue of type I error rate: null hypothesis is erroneously rejected under scenarios of any $\theta_k =0$, setting $\delta_U=0.25$ and $\zeta = 0.975$.}
    \label{tab:typeI}
    \begin{tabular}{ll|cccccc|c} 
    % <-- Alignments: 1st column left, 2nd middle and 3rd right, with vertical lines in between
      & & \multicolumn{6}{c|}{Subtrial}  &  \\
       \cline{3-8}
      & & 1 & 2 & 3 & 4 & 5 & 6   & {\bf Overall} \\
      \hline
\textbf{Scenario 7}   & Standard HM & 0.0000  & 0.0000  & 0.0000  & 0.0000  & - & -  & 0.0000 \\
& No borrowing  & 0.0036  & 0.0077  & 0.0056  & 0.0100  & - & -   & 0.0269  \\
& EXNEX & 0.0001  & 0.0003  & 0.0002  & 0.0000   & - & -  & 0.0006 \\
& Proposed approach & 0.0073 & 0.0089 & 0.0002 & 0.0002 & - & -  & 0.0166 \\
\hline
\textbf{Scenario 8}   & Standard HM  & - & 0.0155 & -  & -  & 0.0080 & -  & 0.0207 \\
& No borrowing   & - & 0.0077 & - & -  & 0.0008 & -  & 0.0085  \\
& EXNEX     & - & 0.0195 & - & -  & 0.0056 & -   & 0.0251 \\
& Proposed approach   & - & 0.0155 & - & - & 0.0017 & -  & 0.0172  \\
      \hline
{\textbf{Scenario 9}}      & Standard HM & 0.0000   & 0.0000   & 0.0000   & 0.0000   & 0.0000   & 0.0000     & 0.0000 \\
      & No borrowing & 0.0036 & 0.0077 & 0.0056 & 0.0100 & 0.0008 & 0.0006 & 0.0283 \\
      & EXNEX  & 0.0000 & 0.0001 & 0.0000 & 0.0000 & 0.0000 & 0.0000  & 0.0001  \\
      & Proposed approach & 0.0014  & 0.0028  & 0.0000  & 0.0000  & 0.0002  & 0.0020   & 0.0064  \\
\bottomrule
    \end{tabular}
  \end{center}
\vspace{-0.8em}
* {\bf Overall}: the proportion of trials with erroneous {\em Go} decision for at least one subtrial.
\end{table}

\clearpage



\section*{Supplementary material}

# Web-based Supplementary Materials for:
Borrowing of information across patient subgroups in a basket trial based on distributional discrepancy

**by Haiyan Zheng, James M.S. Wason**

**A. TRANSFORMATION OF PAIRWISE DISTANCES INTO PROBABILITY WEIGHTS**

In the main manuscript, we used Equation (3.11) to transform the pairwise Hellinger distances into probability weights, which contribute to combining the commensurate predictive priors specified based on subtrial data $x_k$ complementary to subtrial $k^\star$ of our analysis interest. In this section, we present the property of this transformation analytically. How different values of $s_0$ may impact the trial operating characteristics is illustrated numerically with simulation results in Section C.

**Property 1**: Given a vector of equal pairwise $d_{kk^\star}$, equal weights $p_{kk^\star} = \frac{1}{K-1}$ will be obtained for any $s_0 > 0$.

**Property 2**: A large value of $s_0 \gg 1 \geq d_{kk^\star}$ yields about equal weights $p_{kk^\star} \to \frac{1}{K-1}$ irrespective of the magnitude of differences between $d_{kk^\star}$.

**Proof**: Suppose all the pairwise $d_{kk^\star}$ are identical to $0 \leq d \leq 1$. Following Equation (3.11), we have

$$p_{kk^\star} = \frac{\exp(-d/s_0)}{\underbrace{\exp(-d/s_0) + \cdots + \exp(-d/s_0)}_{(K-1) \text{ terms}}} = \frac{1}{K-1}.$$

We constrain $s_0 > 0$ so as to have a decreasing function of the distance measure. On the other hand, when $s_0 \gg 1 \geq d_{kk^\star}$, $\exp(-d_{kk^\star}/s_0) \to \exp(0) = 1$. Therefore, $p_{kk^\star} \to \frac{1}{K-1}$.

**Proposition**: The difference between weights $p_{kk^\star}$ transformed from $0 \leq d_{kk^\star} \leq 1$ decreases, as the value of $s_0$ increases.

**Proof**: Let $d_{tk^\star}$ and $d_{jk^\star}$ be any two pairwise Hellinger distances from the $k^\star$-th column of the $K \times K$ distance matrix, and we further suppose that $d_{tk^\star} \leq d_{jk^\star}$. Let $p_{tk^\star}$ and $p_{jk^\star}$ be the weights obtained following Equation (3.11), respectively, in the presence of other pairwise distances to $\theta_{k^\star}$. Denoting the difference by $\Delta_p = p_{tk^\star} - p_{jk^\star}$, we derive the derivative of $\Delta_p$ with respect to $s_0$ to illustrate how $\Delta_p$ changes (increase or decrease), as $s_0$ increases.

$$\begin{aligned}
\Delta_p &= p_{tk^\star} - p_{jk^\star} \\
&= \frac{\exp(-d_{tk^\star}/s_0)}{\sum_k \exp(-d_{kk^\star}/s_0)} - \frac{\exp(-d_{jk^\star}/s_0)}{\sum_k \exp(-d_{kk^\star}/s_0)} \\
&= \frac{\exp(-d_{tk^\star}/s_0) - \exp(-d_{jk^\star}/s_0)}{\sum_k \exp(-d_{kk^\star}/s_0)}
\end{aligned} \quad (1)$$

Let $f = \exp(-d_{tk^\star}/s_0) - \exp(-d_{jk^\star}/s_0)$ and $g = \sum_k \exp(-d_{kk^\star}/s_0)$. Following the *quotient rule*, we know that $\Delta_p' = \frac{f'g - fg'}{g^2}$. The interest is to learn the rate of change of $\Delta_p$ changes as the value of $s_0$ varies, through the derivative. In the following, we obtain $f'$ and $g'$, respectively.

1

$$f' = \left\{\exp\left(-\frac{d_{tk^\star}}{s_0}\right)\right\}' - \left\{\exp\left(-\frac{d_{jk^\star}}{s_0}\right)\right\}' \quad (2)$$
$$= \frac{1}{s_0^2} d_{tk^\star} \exp\left(-\frac{d_{tk^\star}}{s_0}\right) - \frac{1}{s_0^2} d_{jk^\star} \exp\left(-\frac{d_{jk^\star}}{s_0}\right),$$

and

$$\begin{aligned}
g' &= \left[\sum \exp\left(-\frac{d_{kk^\star}}{s_0}\right)\right]' \\
&= \left[\exp\left(-\frac{d_{1k^\star}}{s_0}\right) + \exp\left(-\frac{d_{2k^\star}}{s_0}\right) + \cdots + \exp\left(-\frac{d_{Kk^\star}}{s_0}\right)\right]' \\
&= \exp\left(-\frac{d_{kk^\star}}{s_0}\right) \cdot (-d_{1k^\star}) \cdot (-1)s_0^{-2} + \cdots + \exp\left(-\frac{d_{Kk^\star}}{s_0}\right) \cdot (-d_{Kk^\star}) \cdot (-1)s_0^{-2} \\
&= s_0^{-2} \left[d_{1k^\star} \exp\left(-\frac{d_{1k^\star}}{s_0}\right) + \cdots + d_{Kk^\star} \exp\left(-\frac{d_{Kk^\star}}{s_0}\right)\right].
\end{aligned} \quad (3)$$

Thus,

$$\begin{aligned}
f'g - fg' &= \left[\frac{1}{s_0^2} d_{tk^\star} \exp\left(-\frac{d_{tk^\star}}{s_0}\right) - \frac{1}{s_0^2} d_{jk^\star} \exp\left(-\frac{d_{jk^\star}}{s_0}\right)\right]\left[\exp\left(-\frac{d_{1k^\star}}{s_0}\right) + \cdots + \exp\left(-\frac{d_{Kk^\star}}{s_0}\right)\right] \\
&\quad - \left[\frac{1}{s_0^2}\exp\left(-\frac{d_{tk^\star}}{s_0}\right) - \frac{1}{s_0^2}\exp\left(-\frac{d_{jk^\star}}{s_0}\right)\right]\left[d_{1k^\star}\exp\left(-\frac{d_{1k^\star}}{s_0}\right) + \cdots + d_{Kk^\star}\exp\left(-\frac{d_{Kk^\star}}{s_0}\right)\right] \\
&\leq \left[\frac{1}{s_0^2} d_{tk^\star} \exp\left(-\frac{d_{tk^\star}}{s_0}\right) - \frac{1}{s_0^2} d_{jk^\star} \exp\left(-\frac{d_{jk^\star}}{s_0}\right)\right]\left[\exp\left(-\frac{d_{1k^\star}}{s_0}\right) + \cdots + \exp\left(-\frac{d_{Kk^\star}}{s_0}\right)\right] \\
&\quad - \left[\frac{1}{s_0^2}\exp\left(-\frac{d_{tk^\star}}{s_0}\right) - \frac{1}{s_0^2}\exp\left(-\frac{d_{jk^\star}}{s_0}\right)\right]\left[\exp\left(-\frac{d_{1k^\star}}{s_0}\right) + \cdots + \exp\left(-\frac{d_{Kk^\star}}{s_0}\right)\right] \\
&= \frac{1}{s_0^2}\left[\exp\left(-\frac{d_{1k^\star}}{s_0}\right) + \cdots + \exp\left(-\frac{d_{Kk^\star}}{s_0}\right)\right]\left[\exp\left(-\frac{d_{tk^\star}}{s_0}\right)(d_{tk^\star}-1) - \exp\left(-\frac{d_{jk^\star}}{s_0}\right)(d_{jk^\star}-1)\right] \leq 0
\end{aligned} \quad (4)$$

The last step of the inequality follows, because $h(d) = \exp(-\frac{d}{s})(d-1)$ is monotonically increasing as $d$ increases, at a given value of $s$.

Therefore, $\Delta'_p = \frac{f'g - fg'}{g^2} \leq 0$, which means $\Delta_p$ is decreasing for all $s_0$ in the interior of $\mathbb{S} = (0, \infty)$. It suggests that a smaller value of $s_0$ discerns the differences between $d_{kk^\star}$ more sensitively. This Proposition can also be linked to Property 2 that we argue against setting $s_0 \gg 1 \geq d_{kk^\star}$.

## B. SPECIFYING THE SPIKE-AND-SLAB PRIOR

We place a spike-and-slab prior on the precision factor of the predictive prior, $\nu$. Data from an incommensurate subtrial are down-weighted through the 'slab' prior Unif$(\mathcal{B}_1, \mathcal{B}_2)$, from which $\frac{1}{\nu}$ has an inverse uniform distribution with the probability density function as $g_k(x) = (\mathcal{B}_2 - \mathcal{B}_1)^{-1} x^{-2}$ and the support on $[\mathcal{B}_2^{-1}, \mathcal{B}_1^{-1}]$. The point-to-point borrowing of information from a specific subtrial for commensurability is through the point-mass prior, $\mathcal{S}$, so-called the 'spike'.

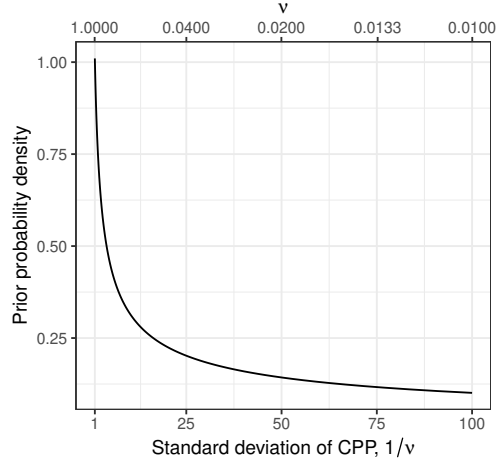

Figure S1: Prior pobability density of the standard deviation of a commensurate predictive prior (CPP) under the slab prior for down-weighting.

Following the stipulation (3.3) of the main manuscript, information on the treatment effect in subtrial $k$ is represented into a predictive distribution, with the normal precision $\nu_{kk^\star}^2$ which determines the degree of borrowing. In the simulation study, we set $\mathcal{B}_1 = 0.01$, $\mathcal{B}_2 = 1$ and $\mathcal{S} = 100$ to define the spike-and-slab for each $\nu_{kk^\star}$ for the strong down-weighting or borrowing of information in the extreme cases of prior probability $w_{kk^\star} = 1$ or 0, respectively. Relating this specification of slab-and-spike prior to the parameter configurations of generating trial data, where the standard deviation (SD) for $\theta_k$ is expected to be smaller than 0.4 (we set the SD of patient responses $y_{ik}$ as $\sigma = 0.4$). Figure S1 visualises the prior probability density distribution of the 'slab' prior used for simulations, from which we know with probability of 96.0%, the SD of the commensurate predictive prior $1/\nu_{kk^\star} \in [1, 25]$, and 98.0% that $1/\nu_{kk^\star} \in [1, 50]$. Even when $1/\nu_{kk^\star} = 1$ at the bound, this is sufficient to substantially discount the information from subtrial $k$. The 'spike' prior was specified so that $x_k$ can be *completely pooled* into subtrial $k^\star$ for perfect data consistency, which can be suggested by a $d_{kk^\star} = 0$. In such situations, the proposed methodology regards patients of the two subtrials as exchangeable irrespective of the subgroup label.

## C. ADDITIONAL SIMULATION RESULTS

In this section, we report and interpret additional simulation results for comparing the proposed methodology with several alternative Bayesian models.

Figure S2 shows the median width of the 95% posterior credible intervals (CIs) for $\theta_k$ using bars of different heights, together with the 10th and 90th percentiles using endpoints of the error bars. This help illustrate how precise the Bayesian models can give the posterior estimates, in addition to the bias and MSE comparisons in Section 4 (Figure 1) of the main manuscript. As we can see, the approach of no borrowing leads to the widest posterior CIs across all scenarios. In contrast, the proposed approach (borrowing based on distributional discrepancy) has the narrowest posterior CIs, when there exist at least one subtrial with similar treatment effect. In scenarios 5 and 6, where data are consistent across subtrials, the proposed approach outperforms the other Bayesian models by producing the narrowest posterior CIs. Looking at, for example, subtrials 3 and 4 in scenario 1 that have relatively moderate treatment effect in a basket trial, the proposed approach leads to posterior estimates with the CI widths comparable to those of the standard HM or EXNEX,

3

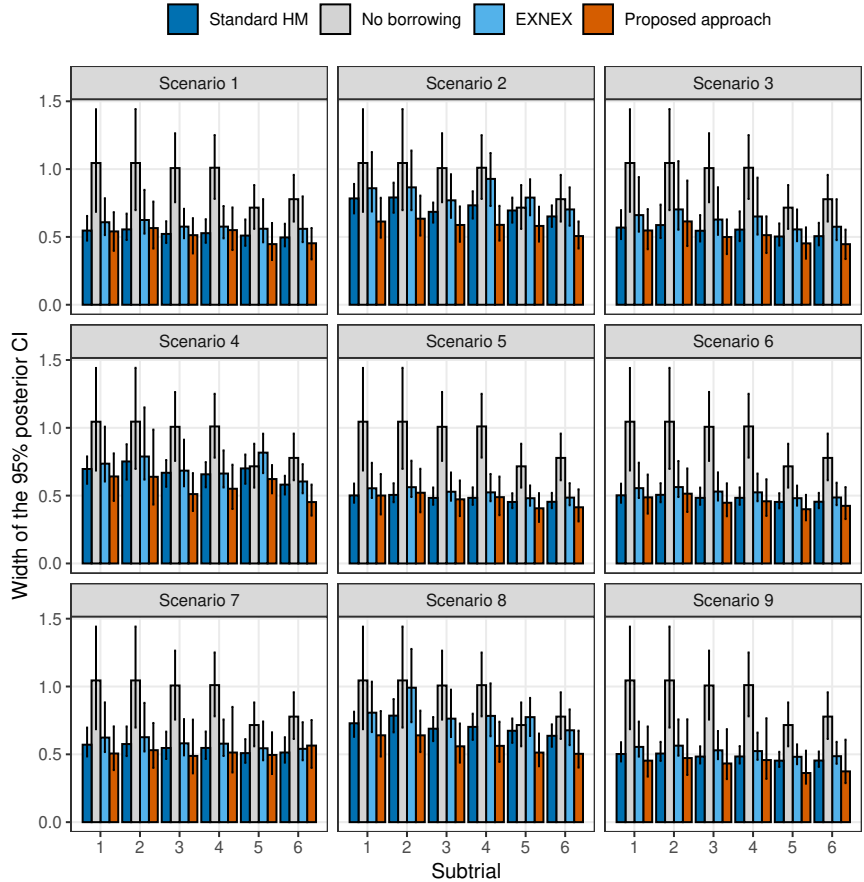

Figure S2: A bar chart for the width of the credible interval, bounded by the 2th and 97.5th percentiles, of the posterior distribution for $\theta_k$, $k = 1, \ldots, 6$ estimated using different Bayesian analysis models.

but much wider error bars. This suggests that our approach using a discrepancy measure is more sensitive to the disparate treatment effect on patient subgroups demonstrated by trial data. By identifying subtrials with similar treatment effect, it can ensure the inference for a subtrial with an extreme treatment effect with information leveraged from subtrials with the most commensurate treatment effect or no borrowing at all. Whereas, the alternative approaches of borrowing (such as the standard HM) tend to shrink the subgroup-specific estimates towards the overall population mean, which is often close to the medium effect size as an average across all subtrials.

In Section A, we demonstrated two essential properties of the transformation of the pairwise Hellinger distances into weights to identify the most commensurate subtrial(s). We now present numerically how the proposed methodology differentiates the degree of borrowing from other subtrials, which may have either commensurate or incommensurate treatment effects. Specifically, it is desirable to allocate the largest weight(s) to the most consistent complementary subtrial(s) in a scenario of heterogeneous treatment effects, but nearly equal weights in a scenario of subtrials with consistent treatment effects.

Based on 5000 simulation replicates, Figure S3 reports the weight allocation to the $(K-1)$ commensurate predictive priors, as a function of the computed pairwise commensurability, for scenarios 4 (some subtrials are more commensurate between themselves than with others) and 5 (equally commensurate between all). The subfigures show our approach can correctly identify the most consistent subtrials. Referring to subtrial 5 with the lowest treatment effect of subfigure

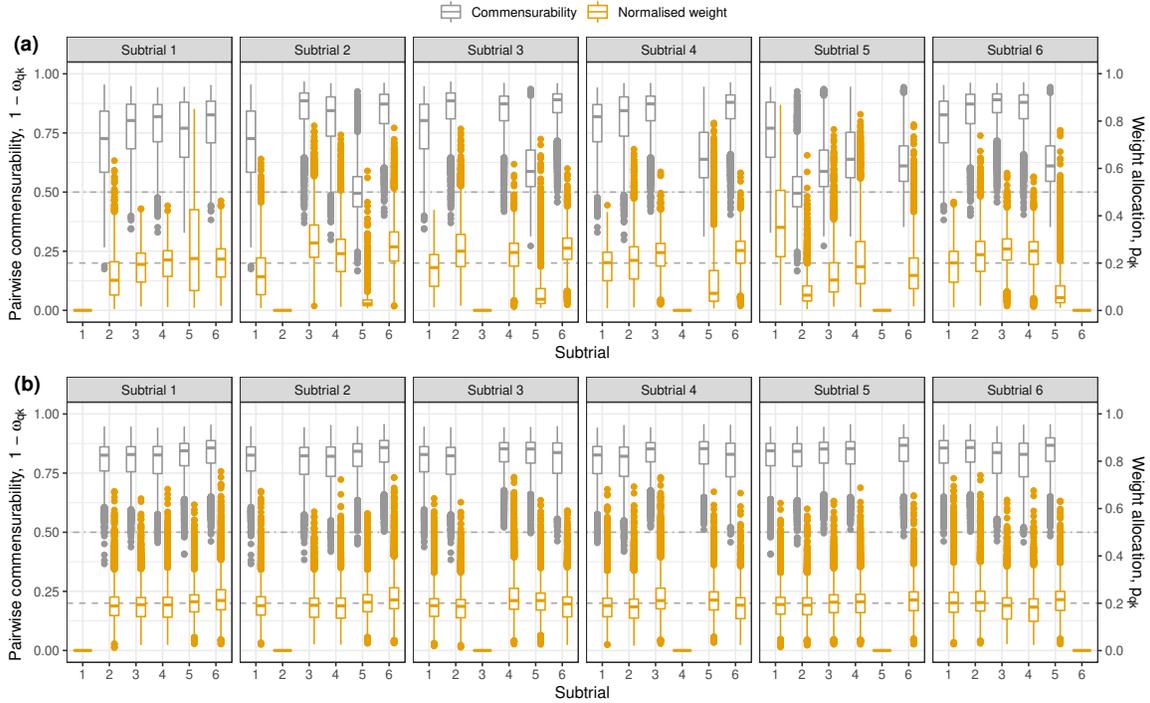

Figure S3: Boxplots of the weight allocation to a specific subtrial ($q \neq k$) for leveraging information into the subtrial $k$ of current analysis interest. Subfigures (a) and (b) visualise the simulation results of scenarios 4 and 5, repsectively. The dashed horizontal lines indicate the level of commensurability as 0.5, and the weight of 0.2 which is the case when all subtrials are equally commensurate.

(a), for example, we read the largest weight is allocated to subtrial 1 and the smallest to subtrial 2, with the medians as 0.351 and 0.06, respectively. Subfigure (b) shows that equal weights are allocated to the other subtrials in a case of all subtrials with the same treatment effect. The median weights are 0.2 for the other subtrials respectively, as is illustrated in Section A. The interquartile range of each boxplot, for pairwise commensurability and correspondingly the weight, tend to be large for subtrials that have a medium size of treatment effect. For example, within the same plot for subtrials 2 – 6 of subfigure (a), the interquartile range of the boxplot of subtrial 1 is larger than any other subtrials to the same target subtrial. Fitting a linear regression model to the simulated basket trial data, $\theta_1$ may be estimated as larger (smaller) than the true $\theta_1$ and regarded to be more commensurate to subtrials with a high (low) treatment effect, such as subtrial 2 (5).

We move onto the exploration of how the specification of $s_0$ would impact the sensitivity for identifying the most commensurate subtrial(s). Figure S4 presents boxplots of the weights, which are transformed from the same pairwise Hellinger distances, to reflect the relative importance of other subtrials in scenario 4, where we have set $s_0 = 0.15$ (used for the main simulation study), 0.25, 0.35 and 0.45. Taking the plot for Subtrial 5 in Figure S4 as an example, we observe that as $s_0$ increases, the weights allocated to other subtrials (despite the different magnitudes of incommensurability) converge to the value of 0.2. When nearly equal weights are yielded, it means that all the other subtrials are regarded as equally (in)commensurate. Robust inferences can still be achieved, as the individual point-to-point commensurate priors are determined based on the computed pairwise Hellinger distance for appropriate amount of borrowing. But setting a large value of $s_0$ leads the proposed methodology to lose the ability of identifying the commensurate subtrials sensitively. As was demonstrated in Section A, we argue against a $s_0 \gg 1 \geq d_{kk^\star}$ which would result in about equal weights for the other subtrials even if the treatment effects may vary.

5

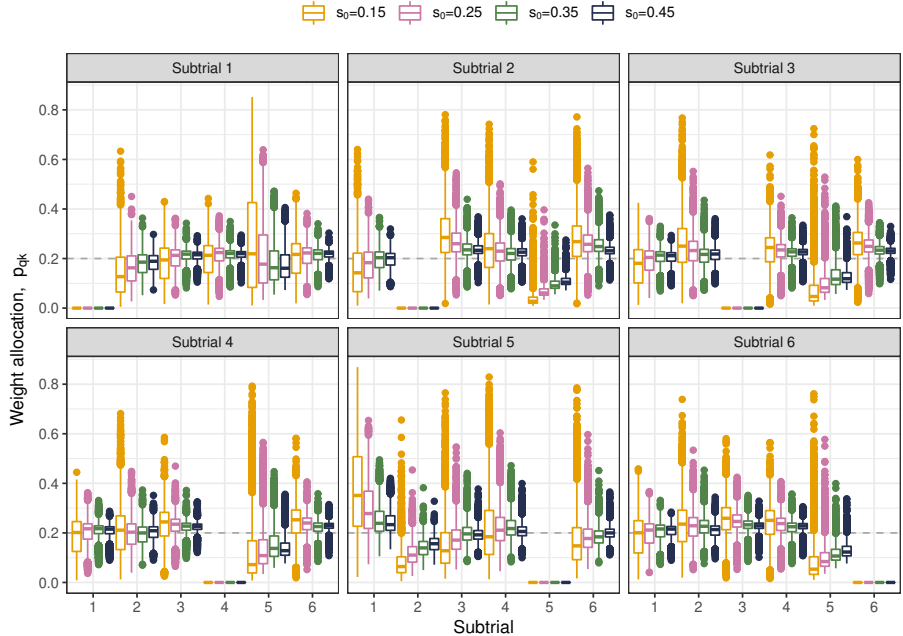

Figure S4: Boxplots of the weight allocation to complementary subtrials, transformed from the pairwise Hellinger distance setting $s_0 = 0.15, 0.25, 0.35, 0.45$ in scenario 4.

Finally, we note that the *Go* or *No-go* decision made for each subtrial depends on the interval probability computed accroding to the criterion defined in Section 2 of the main manuscript. The operating characteristics, i.e., analogues of the statistical power and type I error rate, vary relying on the choices of the threshold, $\delta_U$, and the level $\zeta$. In Section 4 of the main manuscript, we have reported a subset of the simulation results, setting $\delta_U = 0.25$ and $\zeta = 0.975$. In Figure S5 and Table S1 below, we present the complementary simulation results of operating characteristics produced from $\delta_U = 0.30$ and $\zeta = 0.975$.

Looking across Figure 2 ($\delta_U = 0.25$) of the main manuscript and Figure S5 ($\delta_U = 0.35$), we observe that the Bayesian models yield lower power in general given a larger threshold $\delta_U$. The magnitude of decrease by different models may not be the same, particularly for the approaches of borrowing, where the power curve is a function of both the treatment effect and the information leveraged from other subtrials. In particular, the stipulation of the half-normal priors and the 'spike' prior determines the maximum amount of information that can be leveraged. When the threshold $\delta_U$ is levelled up, it becomes more difficult to make a *Go* for a subtrial with $\theta_k = 0$. The analogue of type I error rate recorded in Table S1 is therefore smaller than that is in Table 2 of the main manuscript.

### D. SOME CONSIDERATIONS ON THE SUBTRIAL SAMPLE SIZE

In the main manuscript, we evaluated the proposed Bayesian method in the simulation study for analysing basket trials with unequal subtrial sample sizes. Nevertheless, the subtrial sample sizes, ranging from 10 to 20, do not vary by much. We note there are scenarios when a basket trial would be planned to include rare disease subgroup(s), in which a standard phase II trial may be unrealistic to be undertaken in light of the scant sample size. It would therefore be interesting to see how the proposed method works in those cases, as pointed out by a reviewer. We are

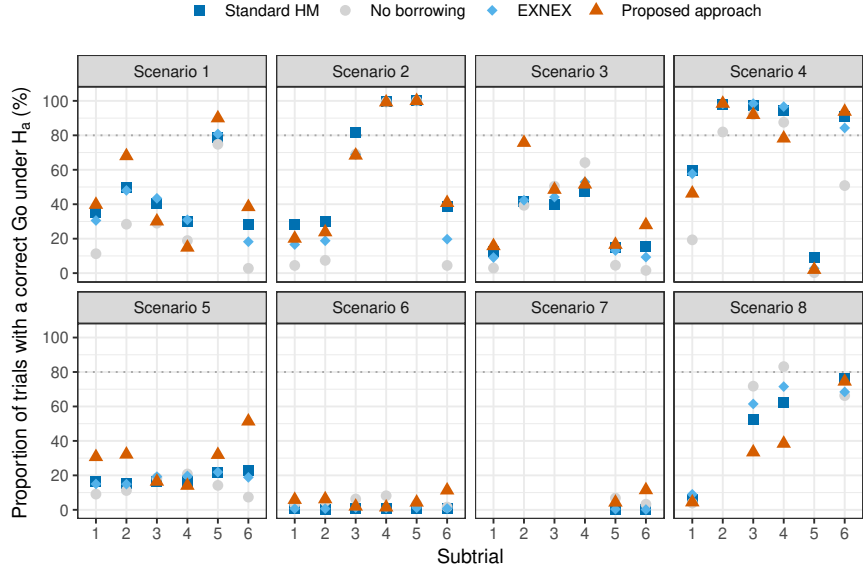

Figure S5: Comparison of the Bayesian analysis models with respect to the analogue of statistical power: null hypothesis is correctly rejected in the presence of a treatment effect per subtrial, setting $\delta_U = 0.30$ and $\zeta = 0.975$.

Table S1: Comparison of the Bayesian analysis models with respect to the analogue of type I error rate: null hypothesis is erroneously rejected under scenarios of any $\theta_k = 0$, setting $\delta_U = 0.30$ and $\zeta = 0.975$.

|  |  | Subtrial | | | | | | Overall |
|---|---|---|---|---|---|---|---|---|
|  |  | 1 | 2 | 3 | 4 | 5 | 6 |  |
| **Scenario 7** | Standard HM | 0.0000 | 0.0000 | 0.0000 | 0.0000 | - | - | 0.0000 |
|  | No borrowing | 0.0025 | 0.0052 | 0.0035 | 0.0063 | - | - | 0.0175 |
|  | EXNEX | 0.0000 | 0.0000 | 0.0000 | 0.0000 | - | - | 0.0000 |
|  | Proposed approach | 0.0033 | 0.0055 | 0.0001 | 0.0000 | - | - | 0.0089 |
| **Scenario 8** | Standard HM | - | 0.0058 | - | - | 0.0026 | - | 0.0084 |
|  | No borrowing | - | 0.0052 | - | - | 0.0004 | - | 0.0056 |
|  | EXNEX | - | 0.0100 | - | - | 0.0022 | - | 0.0122 |
|  | Proposed approach | - | 0.0032 | - | - | 0.0000 | - | 0.0032 |
| **Scenario 9** | Standard HM | 0.0000 | 0.0000 | 0.0000 | 0.0000 | 0.0000 | 0.0000 | 0.0000 |
|  | No borrowing | 0.0025 | 0.0052 | 0.0035 | 0.0063 | 0.0004 | 0.0004 | 0.0182 |
|  | EXNEX | 0.0000 | 0.0001 | 0.0000 | 0.0000 | 0.0000 | 0.0000 | 0.0001 |
|  | Proposed approach | 0.0009 | 0.0017 | 0.0000 | 0.0000 | 0.0000 | 0.0008 | 0.0034 |

\* **Overall**: the proportion of trials with erroneous *Go* decision for at least one subtrial.

particularly concerned about whether effective borrowing of information could enhance inference for a small subtrial, when the sufficient statistics (i.e., mean and variance) of the subtrial data are nearly identical apart from the fact that $n_q \ll n_k$, with $q \neq k$. For simplicity, in the hypothetical data examples that we will present and interpret below, we constrain that the basket trials involve two subgroups only, thus $K = 2$.

We simulate the hypothetical basket trial data setting the 'true' parameter value for $\theta_1 = \theta_2 = 0.8$ and following the settings in Section 4.1 of the main manuscript, under two realistic scenarios where $n_1 = n_2 = 20$ and where $n_1 = 20, n_2 = 200$. In particular, the trial structure and the parameters other than $\theta_k$ for generating and analysing the basket trial data remain unchanged.

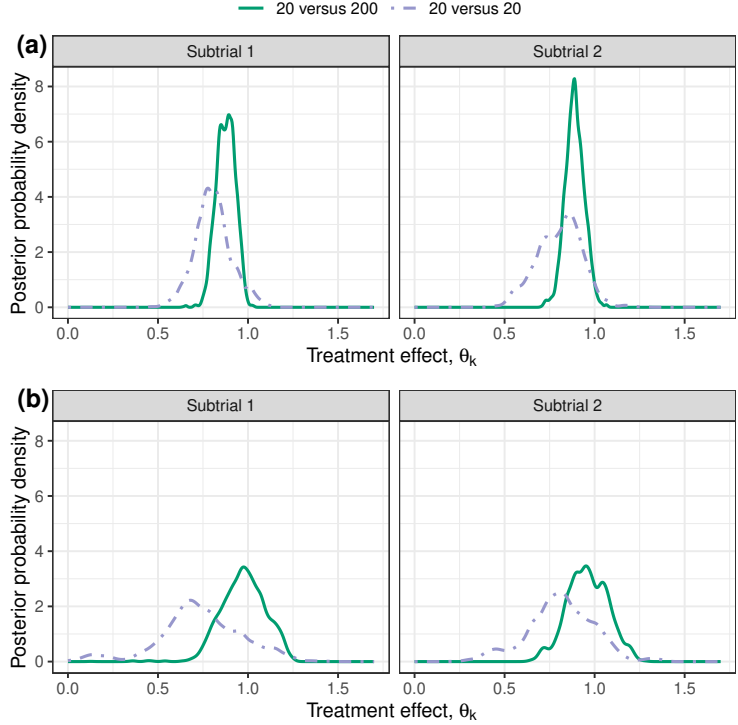

Figure S6: Posterior pobability density of the treatment effect $\theta_k$, with information leveraged from a complementary subtrial with a similar or different sample size. Subfigures (a) and (b) correspond to the analysis results of data simulated by setting the inter-patient standard deviation $\sigma = 0.4$ and $0.8$, respectively.

We present two sets of hypothetical data examples below. In the first set with relatively small inter-patient variability ($\sigma = 0.4$), the operational posteriors are approximately $N(0.808, 0.12^2)$ and $N(0.804, 0.11^2)$ in the scenario of $n_1 = n_2 = 20$, which suggests the subtrial data as highly commensurate. The strong borrowing of information is permitted following our methodology, as evidenced by the Hellinger distance eventually assessed as $d_H(\pi_{\theta_1}, \pi_{\theta_2}) = 0.05$ for the posteriors leveraging complementary subtrial data. For comparison, in the scenario where $n_1 = 20, n_2 = 200$, the operational posteriors are approximately $N(0.762, 0.19^2)$ and $N(0.863, 0.05^2)$, respectively. The Hellinger distance between the posteriors based on all subtrial data is 0.08, which is about the same level of commensurability as that of the scenario of equal subtrial sample size. In the second set of data examples with a larger inter-patient variability ($\sigma = 0.8$), the operational posteriors are approximately $N(0.484, 0.27^2)$ and $N(0.463, 0.28^2)$ in the equal subtrial sample size scenario, and $N(1.277, 0.55^2)$ and $N(0.884, 0.15^2)$ in the scenario of $n_1 = 20, n_2 = 200$. The Hellinger distances between the posteriors based on all subtrial data are both 0.16 for the two sample size scenarios.

Figure S6 visualises the posterior density curves of $\theta_k$ with information leveraged from the complementary subtrial based on the proposed method. Subfigure (a) corresponds to the first hypothetical dataset (with $\sigma = 0.4$) and subfigure (b) to the second (with $\sigma = 0.8$). Sample size scenarios are distinguished by the colour and line type of the pdf curves: gray and dashed for the scenario $n_1 = n_2 = 20$, while green and solid for that of $n_1 = 20, n_2 = 200$.

Here, we would like to draw the reader's attention to the plots labelled 'Subtrial 1' within both subfigures (a) and (b), as the interest centres around the informativeness of $\theta_1$ (the treatment effect in the smaller subtrial) with data leveraged from subtrial 2 (particularly in the scenario of $n_2 = 200$). As reported, the Hellinger distances computed in both scenarios are comparable and

both suggesting high commensurability. How informative the posterior of $\theta_1$ could be depends on (i) the data exclusively from subtrial 1, and (ii) the complementary subtrial data together with the Hellinger distance. Comparing the gray with the green curves (across sample size scenarios) in the Subtrial 1 plots, we observe that incorporating the complementary data with $n_2 = 200$ increases the informativeness of the posterior for $\theta_1$ substantially. It drags the curve shape (height and dispersion) towards that of the green curve shape in the plot labelled 'Subtrial 2' which is largely determined by the subtrial data with $n_2 = 200$ in the same subfigure.

These data examples suggest that the proposed approach can accommodate circumstances when the subtrial sample sizes of a basket trial are imbalanced. Since the Hellinger distance guides the magnitude of borrowing through the distribution parameters (estimated mean and variance of the operational posterior), the complementary subtrial data, regardless of the actual sample size, can be leveraged properly in the considered setup (where the sample size of the small subtrial is not that extreme). However, this area deserves further research for more extreme scenarios with $n_1 \ll n_2$. Additional complexity would arise if the inter-patient variability is different across subtrials. It is beyond the scope of the present paper, but we envisage it could be significant, and also relevant to future basket trials which may adopt asynchronous recruitment. For example, by the time a new subtrial is open to enroll patients, some complementary subtrials have already accrued a good amount of information.



\end{document}